\documentclass[aps,twocolumn,10pt]{revtex4-1}
\usepackage{amsmath}
\usepackage{amssymb}
\usepackage{graphicx}
\usepackage{dcolumn}
\usepackage{bm}
\usepackage[english]{babel}
\usepackage{hyperref}         
\hypersetup{colorlinks=true}  

\makeatletter
\let\reftagform@=\tagform@
\def\tagform@#1{\maketag@@@{\color{blue}(\ignorespaces{#1}\unskip\@@italiccorr)}}
\renewcommand{\eqref}[1]{\textup{\reftagform@{\ref{#1}}}}
\makeatother

\usepackage{color} 
\definecolor{PaleYellow1}{rgb}{1.0,1.0,0.5}
\definecolor{Yellow1}{rgb}{1.0,1.0,0.5}
\definecolor{LinkColor}{rgb}{0.0,0.45,0.0} 
\definecolor{Grey0}{rgb}{0.95,0.95,0.95} 
\definecolor{Grey1}{rgb}{0.92,0.92,0.92} 
\definecolor{Grey2}{rgb}{0.4,0.4,0.4} 
\definecolor{Green1}{rgb}{0.4,0.9,0.4}
\definecolor{Orange1}{rgb}{1.0,0.7,0.05}
\definecolor{DarkBlue}{rgb}{0,0.08,0.45}
\definecolor{DarkRed}{rgb}{0.75,0.08,0.0}
\definecolor{BrightGrey}{rgb}{0.4,0.4,0.4}
\definecolor{BrightGreen}{rgb}{0.1,0.8,0.1}
\definecolor{Orange}{rgb}{1.0,0.5,0.01}
\definecolor{BlueGreen}{RGB}{12,201,179}
\definecolor{kbficolor}{RGB}{171,83,83}
\definecolor{etiscolor}{RGB}{81,106,156}

\newcommand{\B}{\color{DarkBlue}}
\newcommand{\R}{\color{DarkRed}}

\hypersetup{
    colorlinks=true,
    linkcolor=blue,                  
    filecolor=red,   
    urlcolor=blue,                   
    citecolor=LinkColor,             
    anchorcolor=Grey2,
    pdftitle={M.Patriarca - Feynman-Vernon model of a moving thermal environment}, 
    bookmarks=true,
    pdfpagemode=FullScreen,
}

\newcommand{\eq}[1]{Eq.~{\color{blue}(\ref{#1})}}    
\newcommand{\eqs}[1]{Eqs.~{\color{blue}(\ref{#1})}}  

    \makeatletter
    \renewcommand*{\@fnsymbol}[1]{\ensuremath{\ifcase#1
      \or {\color{DarkRed} *}
      \or {\color{DarkRed} **}
      \or {\color{DarkRed} ***}
      \or {\color{DarkRed}\normalsize \textcircled{\footnotesize a}}
      \or {\color{DarkRed}\normalsize \textcircled{\footnotesize b}}
      \or {\color{DarkRed}\normalsize \textcircled{\footnotesize c}}
      \or {\color{DarkRed}\normalsize \textcircled{\footnotesize d}}
      \or {\color{DarkRed}\normalsize \textcircled{\footnotesize e}}
      \or {\color{DarkRed}\normalsize \textcircled{\footnotesize f}}
      \or {\color{DarkRed}\normalsize \textcircled{\footnotesize g}}
      \or {\color{DarkRed}\normalsize \textcircled{\footnotesize h}}
      \or {\color{DarkRed}\normalsize \textcircled{\footnotesize i}}
      \or {\color{DarkRed}\normalsize \textcircled{\footnotesize j}}
      \or {\color{DarkRed}\normalsize \textcircled{\footnotesize k}}
      \or {\color{DarkRed}{3}}
      \or \ddagger
      \or \mathsection
      \or \mathparagraph
      \or \|
      \or **
      \or \dagger\dagger
      \or \ddagger\ddagger
      \else\@ctrerr\fi}}
    \makeatother

\newcommand{\mr}[1]{{\mathrm {#1}}}  
\newcommand{\iu}{\mr{i}}             
\newcommand{\oo}{\omega}             
\newcommand{\oon}{\omega_n}          
\newcommand{\kB}{k_\mr{B}}           
\newcommand{\D}{\mr{D}}              
\newcommand{\bs}{\boldsymbol}              
\renewcommand{\v}{v_\mathrm{env}}              

\begin{document}

\title{Feynman-Vernon model of a moving thermal environment}
\thanks{\small This paper is a revised version with corrections and additional references and figures of the article:\\
Marco Patriarca, {\it Feynman–Vernon model of a moving thermal environment},
Physica E {\bf 29} (2005) 243–250,\\
doi~\href{http://dx.doi.org/10.1016/j.physe.2005.05.021}{{\B {10.1016/j.physe.2005.05.021}}} .
}
\author{Marco Patriarca~}
\thanks{\small 
  Email: {{\tt marco.patriarca}\,@\,{\tt kbfi.ee}}\\
}
\affiliation{\href{www.kbfi.ee}{NICPB--National Institute of Chemical Physics and Biophysics, Tallinn, Estonia.}}
\date{\today}

\begin{abstract}


\noindent
{\bf Abstract}.
This paper reviews the formulation of the Feynman-Vernon model of linear dissipative systems for a standard Brownian particle moving in an external potential $V(x,t)$ and introduces the formulation of a generalized oscillator model of a Brownian particle coupled to a thermal environment moving with a given velocity $\v(t)$.
Diffusion processes in a moving environment are of interest e.g. in the study of the motion of vortices in superfluids.
The starting point of the paper is the formulation of the oscillator model that takes into account space and time invariance of a thermal environment [M. Patriarca, {\it Statistical correlations in the oscillator model of quantum Brownian motion}, Il Nuovo Cimento B, 111(1), 61-72  (1996), {\B doi:~\href{http://link.springer.com/article/10.1007/BF02726201}{{10.1007/BF02726201}}}, \href{https://arxiv.org/abs/1801.02429}{{\R {arXiv:1801.02429}}}],
which has the property of being finite and consistent with the classical limit.
The Langevin equation and the influence functional for a Brownian particle in a moving environment are derived.
\end{abstract}

\maketitle

\nopagebreak
\section{Introduction and Background}

Among the various approaches to quantum Brownian motion explored so far, some are known to be unsuited~\cite{Ray-1979a}, 
while others, such as the category of models based on the oscillator model of a thermal bath, have provided valuable descriptions of that range  of physical phenomena, in which both quantum and statistical fluctuations play a relevant role~\cite{Weiss-2008a,Dattagupta-2014a}.
This category of models includes e.g. the quantum Langevin equation~\cite{Ford1987a,Ford1988a} and  the Feynman and Vernon model~\cite{Feynman1963a,Feynman1965a}, with its later developments by Caldeira and Leggett~\cite{Caldeira1983a,Caldeira1983b}.
The latter model is the subject of the present paper.

\begin{figure}[ht]
 \centering
 \includegraphics[width=7cm]{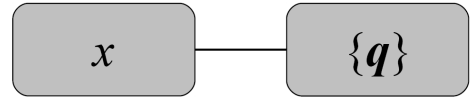}
 \caption{
 \label{scheme}
 In the basic scheme of the oscillator model, the central system $x$ is linearly coupled to the environment oscillators with coordinates $\bs{q} = \{q_1, q_2, \dots\}$.
 A linear coupling does not specify completely the form of total Lagrangian or the initial conditions of the oscillators --- see text for details. 
}
\end{figure}

The possibility to describe dissipation in the quantum regime is due to the fact that the oscillator model can be defined in terms of a Hamiltonian function for the whole system \{central particle + environment\}, analogously to what is done for the quantization of a conservative system.
In the following, for the sake of clarity, a one-dimensional linear dissipative environment is considered, in which an infinite set of harmonic oscillators with coordinates $\{ q_i \} = \{q_1, q_2, \dots\}$ is coupled linearly to the coordinate of the central system $x$, as schematized in Fig.~\ref{scheme}.
Dissipation appears in the system in the limit of a continuous distribution in the angular frequencies $\{\oo_i\}$ of the oscillators, with a suitable density $G(\oo)$, related to the response of the central degree of freedom.

However, as discussed below, the prescription of a linear coupling between bath oscillators and central system does not specify completely the form of the Feynman-Vernon model.
In fact, such ambiguities concern both the form of the Lagrangian and the initial conditions of the bath oscillators and have observable consequences on the classical and quantum dynamics of the central system.

The route to the application of the Feynman-Vernon model to quantum Brownian motion was explored by Caldeira and Leggett, who studied, in particular, the time-dependent problem of a quantum Brownian particle in Refs.~\cite{Caldeira1983a,Caldeira1983b}.
They showed that the white-noise limit of Brownian motion corresponds to an oscillator frequency distribution function
\begin{equation}
\label{wn}
G_{\mathrm{WN}}(\oo) \approx  \frac{ 2 \eta } {\pi m \oo^2} \, ,
\end{equation}
where $m$ is the oscillator mass and $ \eta$ is the friction coefficient appearing in the Langevin equation,
\begin{equation}   \label{LE}
  M\ddot{x}(t)  - f(t) = - \eta \dot{x}(t) + R(t) \,  .
\end{equation}
Here $f(t)$ an external forcing and $R(t)$ the environmental noise force, which is a Gaussian stochastic process defined by the first two moments
\begin{eqnarray}  
  \langle R(t) \rangle = 0 \ , 
  ~~~~
  \langle R(t) R(s) \rangle = 2 \eta \kB T \delta(t - s) \,  , \label{RR2}
\end{eqnarray}
where $T$ is the environment temperature.
The explicit treatment of quantum Brownian motion given by Caldeira and Leggett~\cite{Caldeira1983a} reveals the general ambiguities affecting the Feynman-Vernon model, signalled for example by the appearance of infinite potential terms related to the oscillator sector of the total Lagrangian.
However, even after the removal of these divergences, the model remain affected by further internal inconsistencies when applied e.g. to the study of the motion of a wave packet.
Such additional inconsistencies can arise from the choice of the initial conditions of the bath oscillators, as discussed in Ref.~\cite{Patriarca1996a} and summarized here below.

\begin{figure}[ht]
 \centering
 \includegraphics[width=7cm]{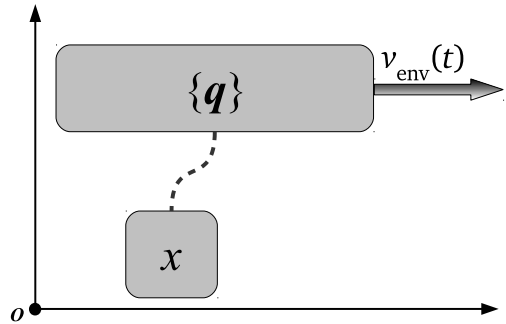}
 \caption{
 \label{scheme-moving}
Scheme of the generalized model with the central system ($x$) coupled to a bath of environment oscillators with coordinates $\bs{q} = \{q_1, q_2, \dots\}$ moving with velocity $\v(t)$. 
}
\end{figure}

The first goal of the present paper is to summarize and discuss the general procedure needed to obtain the central system dynamics avoiding the mentioned problems~\cite{Patriarca1996a}, which consists in\\
(a) properly formulating the Lagrangian of the total system \{central particle\} + \{bath oscillators\};\\
(b) formulating of the corresponding initial conditions of the bath oscillators;\\
(c) finally obtaining the (effective) dissipative dynamics of the central system by integrating the oscillator coordinates.\\
This procedure takes into account some symmetry properties of a standard dissipative environment, namely space translation and reflection invariance (coming from the hypothesis of homogeneity) and time reversal invariance for the initial conditions (following from the assumption of thermal equilibrium).
The reformulation thus obtained provides an intuitive visualization of the environment oscillators and their initial conditions and easily lends itself to be generalized to other types of environments.
For instance, it has been applied to study the motion of a Brownian particle in a constant magnetic field~\cite{2016a-Patriarca} and in thermal environments, which are inhomogeneous in space or time, which can be described through the most general coupling nonlinear in the central particle coordinate but still linear in the environment oscillator coordinates~\cite{Illuminati1994a}.\\
The second goal of the present paper is to illustrate the generality of the procedure by formulating the Feynman-Vernon model for describing a thermal environment moving with an arbitrary velocity $\v(t)$ with respect to the laboratory frame, illustrated in the scheme in Fig.~\ref{scheme-moving}.
Such a type of environments are met in some condensed matter systems and the formalism presented here can be used as a starting point for studying e.g. the effective dynamics of vortices in superconductors, where the superfluid or the normal component of the fluid represents an environment with its characteristic velocity~\cite{Ao1993a,Cataldo-2002a,Kopnin-2002a}.

In Sec.~\ref{classical}, the classical one-dimensional oscillator model of linear dissipative systems is considered and then its generalization to a moving thermal environment is formulated.
The results of the classical problem are then used in Sec.~\ref{quantum} as an effective guiding line for formulating the quantum version, i.e., the corresponding generalized Feynman-Vernon model, summarized in Sec.~\ref{summary}.

\begin{figure*}[ht]
 \centering
 \includegraphics[width=7cm]{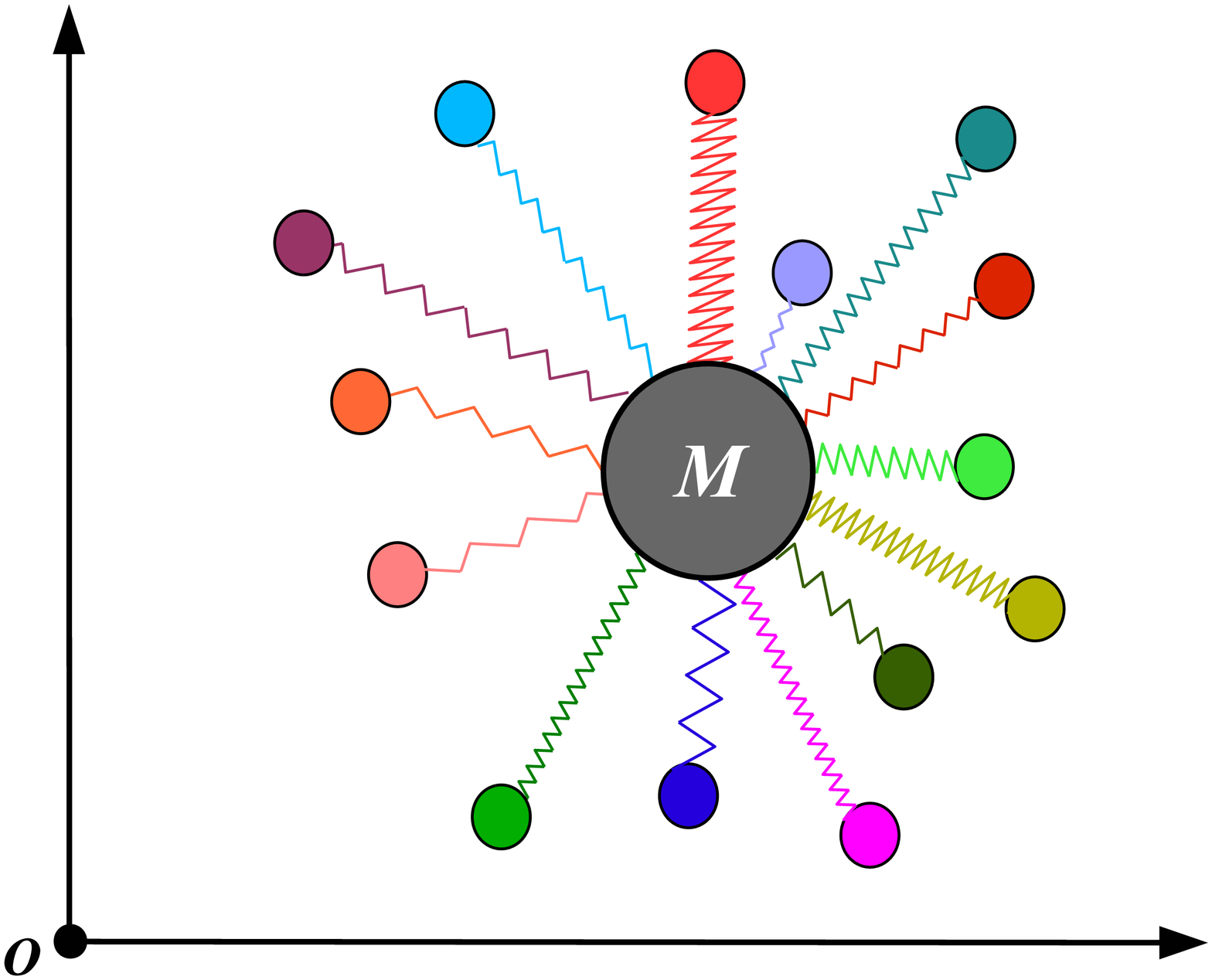}
 \includegraphics[width=7cm]{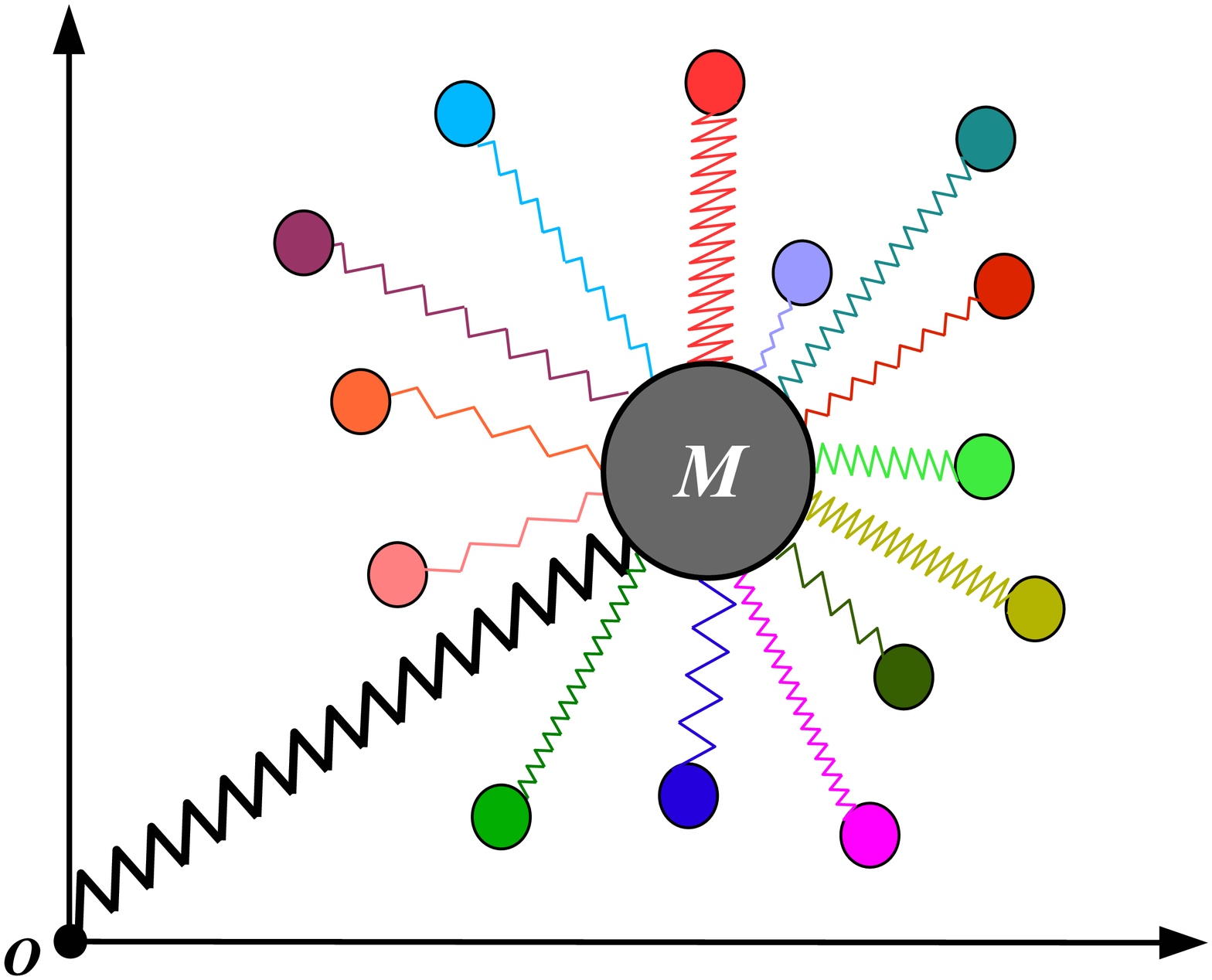}
 \caption{
 \label{model}
 Scheme of the oscillator models corresponding to the Lagrangian in Eq.~(\ref{lagrangian2}) (left) and to that in Eq.~(\ref{lagrangian1})
 for the particular case $c_n = m \omega_n^2$ (right) -- in the latter case there is a spurious attraction toward the origin of the coordinate system.
}
\end{figure*}

\section{Classical Model}
\label{classical}

\subsection{Standard classical model} 

In this section the formulation of the classical version of the oscillator model of dissipative systems is summarized.
As it will appear clear in the following, the classical version contains hints that are crucial for the formulation of the quantum version, discussed in Sec.~\ref{quantum}.
The total system  \{central particle\} $+$ \{environment\} is conveniently described by the Lagrangian
\begin{equation}  \label{Ltot}
  L(x,\dot{x}, \bs{q},\dot{\bs{q}}\})=L_0(x,\dot{x})+\sum_n L_n(q_n,\dot{q}_n,x) \, ,
\end{equation}
where $\bs{q} = \{q_1, q_2, \dots\}$ and $\dot{\bs{q}} = \{\dot{q}_1, \dot{q}_2, \dots\}$ are the sets of the oscillator coordinates and velocities, respectively, $L_0$  the  Lagrangian of the isolated central particle,
and the generic term $L_n$ describes the $n$th environment oscillator and its interaction with the central system.
The limit of a  continuous spectrum can be taken by transforming the sum in \eq{Ltot}
into an integral through a distribution function $G(\oo)$ in the angular frequency, i.e.
\begin{equation} \label{G}
    \sum_n  \phi(\oon) \to \int \delta\oo \, G(\oo) \phi(\oo) \, ,
\end{equation}
for an arbitrary function $\phi(\oo)$.
The Lagrangian $L_0$ is assumed to be given by
\begin{equation} \label{L0}
    L_0(x,\dot{x}) = \frac{1}{2}M\dot{x}^2 - V(x,t) \, ,
\end{equation}
with $V(x,t)$ an external potential.
The Lagrangian $L_0(x,\dot{x})$ of the isolated particle 
leads to Newton's equation of motion, 
$M\ddot{x}  - f = 0$, with $f(x,t) = - \partial V(x,t)/\partial x$.
Therefore the environment sector $\sum_n L_n(q_n,\dot{q}_n,x)$ 
of the total Lagrangian (\ref{Ltot})
must account for the environment (dissipative and random) forces
on the right hand side of \eq{LE} and of their statistical properties in \eq{RR2}, after the elimination of the oscillator variables has been carried out.

The form of \eq{Ltot} suggests a physical interpretation of the generic term $L_n$ as 
the effective Lagrangian for the $n$th oscillator, which also includes
its interaction with the central particle.
In the original perturbative form of 
the oscillator Lagrangian~\cite{Feynman1963a,Feynman1965a},
\begin{equation} \label{lagrangian1}
  L_n(q_n,\dot{q}_n,x) =
  \frac{m}{2}\dot{q}_n^2 + c_n q_n x - \frac{m\oon^2}{2}q_n^2 \, ,
\end{equation}
the interaction strength is determined by the coupling constants $c_n$'s
and one can in principle distinguish between the Lagrangian of the oscillator
and that describing its interaction with the central particle.
However, by using \eq{lagrangian1}, one cannot in general 
reproduce the environment forces in \eq{LE}.
This can be seen e.g. from its transformation properties,
since the dissipative and random forces are assumed to be homogeneous and therefore must be left unchanged 
by a space translation or reflection, on the contrary of \eq{lagrangian1}.
In fact, the interaction with the environment cannot be considered 
as a perturbation to the central particle dynamics, since the environment
is supposed to have relaxed and be in thermal equilibrium in the presence
of the central particle.

In order for the oscillator sector to have the required symmetry properties,
one has first to assume that the oscillator 
and central particle coordinates transform in the same way,
i.e. they all represent physical coordinates.
This is not obvious a priori since no physical meaning for the oscillators
is provided explicitly by the model. 
Translation invariance requires the coupling constants to be given by
$c_n=m\oon^2$, while reflection invariance requires the addition of 
a suitable quadratic potential term to the Lagrangian of each oscillator, so that
the final form of the Lagrangian of the $n$th oscillator 
reads~\cite{Mazur1964a,Ford1965a,Hakim1985a,Schramm1987a,Grabert1988a}
\begin{equation} \label{lagrangian2}
  L_n(q_n,\dot{q}_n,x) = 
  \frac{m}{2} \left[ \dot{q}_n^2 - \oon^2\left(q_n-x\right)^2 \right] \, .
\end{equation}
This Lagrangian suggests a different mechanical picture of the total system,
schematized in Fig.~\ref{model} (left), 
in which  the central particle interacts with an infinite set of oscillators
with their equilibrium position on the central particle itself~\cite{Grabert1988a}.
This is to be contrasted with the picture corresponding to the Lagrangian in Eq.~(\ref{lagrangian1}),
see Fig.~\ref{model} (right), in which spurious attractive forces are present.
More importantly, the non-separability of $L_n(q_n,\dot{q}_n,x)$, according to a perturbative scheme, is to be noticed,
being a natural consequence of the central particle coordinate representing its equilibrium position of the oscillator.

The general link between Lagrangian and initial conditions is in the fact that the form of the Lagrangian $L_n(q_n,\dot{q}_n,x)$ of the $n$th oscillator can be used to determine the corresponding  initial conditions.
Starting from the total Lagrangian (\ref{Ltot}), one can obtain the total Hamiltonian of the system through the Legendre transformation,
\begin{eqnarray}  \label{Htot}
& H(x,p,\bs{q},\bs{p})  &=  \dot{x} p +   \sum_n \dot{q}_n p_n  -  L = \nonumber \\
&                               &= H_0(x,p) + \sum_n H_n(q_n,p_n,x) \, .
\end{eqnarray}
Here $p = \partial L / \partial \dot{x} = M \dot{x}$ and $p_n = \partial L_n / \partial \dot{q_n} = m \dot{q}_n$ are the momenta of the central particle and of the $n$th oscillator, respectively, $\bs{p} = \{p_1, p_2, \dots\}$ the set of the oscillator momenta, $H_0$ the  Hamiltonian of the central particle,
\begin{equation} \label{H0}
    H_0(x,p) = \dot{x} p - L_0 = \frac{p^2}{2 M} + V(x,t) \, ,
\end{equation}
and $H_n$ can be interpreted as the Hamiltonian of the $n$th oscillator,
\begin{equation} \label{hamiltonian1}
  H_n(q_n, p_n, x) = \dot{q}_n p_n - L_n = \frac{p_n^2}{2m} + \frac{m\oon^2}{2} \left(q_n-x\right)^2 \, .
\end{equation}
This is a basic point for the following considerations, since in the oscillator model of dissipative systems it is necessary to assume that the environment is in thermal equilibrium in order to obtain the required statistical properties of the environment forces (fluctuation-dissipation theorem).
Instead, it is possible, from an operative point of view, to perform a measurement of the initial state of the central particle.
In other words, adiabatic initial conditions are assumed, in which the environment oscillators are in thermal equilibrium, relaxing on a much shorter time scale than that characterizing the motion of the central particle, while \emph{no equilibrium is assumed for the central particle}, which may be in principle in an arbitrary initial state.
Correspondingly, the probability distribution $P_0$ of the central particle coordinates is arbitrary, while the corresponding distribution $P_T^{(n)}$ of the generic $n$th oscillator can be written straightforwardly as the canonical  Gibbs-Boltzmann distribution, in the hypothesis that \eq{hamiltonian1} represents the Hamiltonian of the $n$th oscillator:  $P_T^{(n)}(q_n,p_{n}) \propto \exp[ -  H_n(q_n, p_n, x) ] / \kB T$.
Due to the mutual independence of the oscillators, the initial conditions for the total system at the initial time $t=t_0$, going back to configuration space for convenience, are given by the following probability distribution,
\begin{equation} \label{ic}
   P(x_0,\dot{x}_0,\bs{q}_{0},\dot{\bs{q}}_{0})  = P_0(x_0,\dot{x}_0) \prod_n P_T^{(n)}(q_{n0},\dot{q}_{n0}) \, ,
\end{equation}
where $x_0=x(t_0)$, $\dot{x}_0=\dot{x}(t_0)$,  $q_{n0}=q_n(t_0)$, and $\dot{q}_{n0}=\dot{q}_n(t_0)$,  $P_0(x_0,\dot{x}_0)$ is the initial probability density of the central system, and 
\begin{equation} \label{ic1}
   P_T^{(n)}(q,\dot{q})
   = \frac{m\oon}{2\pi\kB T}  \exp\left\{ - {\frac{m}{2{\kB T}} \left[ \dot{q}^2 + \oon^2\left(q-x\right)^2 \right]} \right\} ,
\end{equation}
is the canonical distribution of an oscillator of angular frequency $\oon$.
This normalized canonical distribution of the $n$th oscillator (with a parametric dependence on the central particle coordinate $x$) is unambiguously determined by the oscillator Lagrangian $L_n(q_n,\dot{q}_n,x)$, once the adiabatic hypothesis is assumed.

As an example of initial distribution of the central particle, 
\begin{equation} \label{ic0}
   P_0(x_0,\dot{x}_0) = \delta(x_0 - \bar{x}) \delta(\dot{x}_0 - \bar{v})  ,
\end{equation}
represents the case in which the initial state of the central particle is exactly known, with $\bar{x}$ and $\bar{v}$ representing the values of position and velocity, respectively.
This is an approximation, since in general there will be some uncertainty --- e.g. due to thermal fluctuations --- influencing the measurement process and the initial conditions (\ref{ic0}) should be correspondingly replaced by a distribution with finite $x$- and $v$-width.

Notice that even if the initial probability density in \eq{ic} factorizes into the product of initial probability distributions for each degree of freedom,
it does not factorize into a product of functions of the corresponding coordinates --- thus excluding the possibility of using the so-called ``factorized initial conditions'' frequently employed.

Once the Lagrangian of the total system is known and the initial conditions
assigned, the effective equation of motion for the coordinate $x$
can be easily derived as follows.
The equations of motion for the central particle and the generic $n$th oscillator, obtained from the total Lagrangian defined by \eqs{Ltot} and (\ref{lagrangian2}), are
\begin{eqnarray}
& M\ddot{x} - f(x,t) 					& \, = \, \sum_n m\oon^2(q_n - x) \, , \label{xeq}\\
& m\ddot{q}_n + \oon^2 q_n 	& \, = \, \oon^2 x \, , \label{qneq}
\end{eqnarray}
respectively.
The second equation for the oscillator coordinate $q_n(t)$ can be solved exactly for an arbitrary $x(t)$, with initial conditions $q_n(t_0)=q_{n0}$ and $\dot{q}_n(t_0)=\dot{q}_{n0}$ for the oscillator and $x(t_0) = x_0$ and $\dot{x}(t_0) = \dot{x}_0$ for the central particle.
While the initial central particle coordinates are here assumed to be assigned, according to \eq{ic0},  the oscillator initial coordinates are random variables, consistently with the initial conditions defined by the distribution $P_T^{(n)}(q_{n0},\dot{q}_{n0})$ given in \eq{ic1}.
The solutions $q_n(t)$ can then be replaced in the equation for the central particle, \eq{xeq}.
If \eq{xeq} equation is to be equivalent to the Langevin equation (\ref{LE}), one should be able
to partition the total oscillator force on the right hand side into a systematic -- dissipative -- part depending on the velocity of the central particle and a fluctuating -- random -- force $R(t)$.
This can be done unambiguously by requiring  that the average random force is zero, $\langle R(t) \rangle = 0$.
The result assumes the form of the generalized Langevin equation~\cite{Kubo1957a,Kubo1957b,Kubo1966a,Mori1965a},
\begin{equation}  \label{KM}
  M\ddot{x}(t) - f(x,t) = - \int_{t_0}^{t}d s \, u(t-s)\dot{x}(s) + R(t) ,
\end{equation}
where the correlation function is given by
\begin{equation} \label{u}
  u(\tau) = \sum_n m\oon^2 \cos(\oon\tau) ,
\end{equation}
and the random force is
\begin{equation}  
  \label{R}
  R(t) = \sum_n m\oon^2\left[(q_{n0}-x_0)\cos(\oon t') + \frac{\dot{q}_{n0}}{\oon}\sin(\oon t') \right] ,
\end{equation}
where $t' = t - t_0$.
Using this expressions for $R(t)$ and the initial conditions defined by \eqs{ic} and (\ref{ic1}),
it is straightforward to show that the generalized fluctuation-dissipation theorem holds~\cite{Kubo1957a,Kubo1957b,Kubo1966a,Mori1965a,Weiss-2008a,Dattagupta-2014a},
	\begin{equation}
		\label{FD}
		\langle R(t) \rangle = 0, ~~\langle R(t)R(s) \rangle = \kB T u(t - s) \, .
	\end{equation}
This relation reduces to the white-noise limit of the fluctuation-dissipation theorem, \eq{RR2}, when $G(\oo)$ is given by the white-noise density (\ref{wn}), while correspondingly the friction term in \eq{KM} reduces to the white-noise dissipative force term $-\eta\dot{x}$ of the Langevin equation (\ref{LE}).

\subsection{Generalized classical model} 
\label{gen-class}

Here a generalized model is studied, described by the total Lagrangian in \eq{Ltot}, where in place of \eq{lagrangian2} the following Lagrangian of the $n$th oscillator is used,
\begin{equation} \label{lagrangian3}
  L_n(q_n,\dot{q}_n,x) = 
  \frac{m}{2} \left[ (\dot{q}_n - \v)^2 - \oon^2\left(q_n-x\right)^2 \right] \, ,
\end{equation}
obtained by shifting the oscillator velocities by a common function of time $v_{\rm env}(t)$.
As shown below, $v_{\rm env}(t)$ represents the velocity of the thermal environment, as schematized in Fig.~\ref{scheme-moving}.
Such a picture of a moving thermal environment is valid in the hypothesis that the motion of the environment does not perturb appreciably its state of thermal equilibrium at temperature $T$ (to this aim the acceleration $\dot{v}_{\rm env}(t)$ should be small enough).
The study of the generalized model defined by \eqs{Ltot} and (\ref{lagrangian3}) proceeds in
a way very similar to that of the standard model illustrated above.

The environment is assumed to be in thermal equilibrium
-- and now it is to be assumed also that it relaxes on a time scale 
much shorter than that defined by the function $v_{\rm env}(t)$.
The main difference is in the conjugate momentum of the $n$th oscillator, which now reads $p_n = \partial L_n / \partial \dot{q_n} = m \dot{q}_n - \v$.
Correspondingly, the canonical distribution for the $n$th oscillator,
$P_T^{(n)}(q_n,p_{n}) \propto \exp[-  H_n(q_n,p_n)] / \kB T$,
rewritten in configuration space, reads
\begin{eqnarray}
    P_T^{(n)}(q_{n0},\dot{q}_{n0}) 
    = \frac{m\oon}{2\pi\kB T} \exp\bigg\{&-&\frac{m[\dot{q}_{n0} - \v(t_0)]^2}{2\kB T}~~~~~~
    \nonumber \\
    &-& \frac{m\oon^2 (q_{n0} - x_0)^2}{2\kB T}\bigg\}.
    \label{ic3}
\end{eqnarray}
The Langevin equation for the central particle can be obtained by the same procedure 
used above for the basic model.
While the \eq{xeq} for the central particle is unchanged,
the equation of motion for the $n$th oscillator (\ref{qneq}) is now replaced by
\begin{equation}
\ddot{q}_n + \oon^2 q_n = \oon^2 x + \dot{v}_\mr{env} \, .
\end{equation}
Solving this equation and replacing it in the equation for the central particle now gives
the following generalized Langevin equation,
\begin{equation}  \label{KM1}
  M\ddot{x}(t) - f(x,t) = - \int_{t_0}^{t}d s \, u(t-s)[\dot{x}(s)-\v(s)] + R(t) \, ,
\end{equation}
where $u(\tau)$ is the same correlation function in \eq{u} while 
\begin{eqnarray}
    R(t) = \sum_n m\oon^2	& \bigg[	& (q_{n0}-x_0)\cos(\oon t') \nonumber \\
											& +			& \frac{\dot{q}_{n0} - \v(t_0)}{\oon}\sin(\oon t') \bigg] \, ,
    \label{R2}
\end{eqnarray}
where $t'=t-t_0$.
Notice that both the dissipative and the fluctuating forces now depend on the difference $(\dot{x}-v_{\rm env})$, 
suggesting that $v_{\rm env}$ is the velocity of the thermal environment -- in fact there is no dissipation only when $\dot{x}(t) = v_{\rm env}$.
This ensures that, as in the standard oscillator model, the random force fulfills the fluctuation-dissipation theorem, \eq{FD}, reducing to the white-noise limit (\ref{RR2}) for a frequency density equal to the white-noise frequency density (\ref{wn}).

\section{Quantum model}
\label{quantum}

The classical model illustrated above can be used as a guiding line for the formulation of the quantum model, discussed in this section,
apart from a relevant point concerning the initial conditions of the density matrices of the bath oscillators.

\subsection{ Standard quantum model } 

Here the quantum formulation of the model is summarized~\cite{Feynman1963a,Feynman1965a,Caldeira1983a}.
The total system can be described by a density matrix $ \rho(x,x',\bs{q},\bs{q}',t)$, from which one obtains the reduced density matrix of the central particle by integrating out the environment degrees of freedom,
%
\begin{equation}  \label{rho}
  \rho(x,x',t) = \int dq_1 \int dq_2 \dots \, [\rho(x,x',\bs{q},\bs{q}',t)]_{\bs{q}=\bs{q}'} \, .
\end{equation}
%
The time evolution of the reduced density matrix between two generic times $t_a$ and $t_b > t_a$ can be written as
\begin{equation}  \label{rhob}
  \rho(x_b,x_b',t_b) \!=\!\!  
  \int \! d x_a \, d x_a' \, J(x_b,x_b',t_b|x_a,x_a',t_a) \rho(x_a,x_a',t_a),
\end{equation}
if the initial density matrix of the total system can be factorized analogously to the classical initial conditions (\ref{ic}), i.e.,
\begin{equation} \label{rhod}
    \rho(x_a, x_a', \bs{q}_{a}, \bs{q}_{a}',t_a)
    = \rho_0(x_a,x_a',t_a) \prod_n \rho_n(q_{na},q_{na}') \, ,
\end{equation}
where $\rho_0(x_a,x_a',t_a)$ and $\rho_n(q_{na},q_{na}')$ represent 
the initial density matrix for the central particle and the generic
$n$th oscillator, respectively.
The effective propagator $J(x_b,x_b',t_b|x_a,x_a',t_a)$ in \eq{rhob} has the following path-integral expression,
\begin{eqnarray} \label{J1}
&&	J(x_b,x_b',t_b|x_a,x_a',t_a) \nonumber \\ 
&&	= \int_a^b\D x \, \D x' 
		\exp\left\{\frac{\iu}{\hbar} \left( S_0[x] - S_0[x'] \right) \right\} F[x,x']  ,
\end{eqnarray}
where $a$ is a short notation for the boundary conditions at $t = t_a$, i.e., $x(t_a)=x_a$ and $x'(t_a) = x_a'$;
$b$ represents the analogous conditions at $t = t_b$;
the functional $S_0[x]=\int d t L_0(x,\dot{x})$, with $L_0$ given in \eq{L0}, is the action of the isolated central particle;
and $F[x,x']$ is the influence functional containing the effects of the environment. 
It can be shown that the influence functional factorizes as
\begin{equation}
    F[x,x'] = \prod_n F_n[x,x'] \, ,
\end{equation}
into the product of the contributions  $F_n[x,x']$  of the single oscillators.
The  explicitly form of the influence functional of the $n$th oscillator is given by
\begin{eqnarray}
    F_n[x,x'] 
    &=& \int d q_{nb} \, d q_{na} \, d q_{na}' 
        K_n^*(q_{nb}, t_b | q_{na}', t_a; [x']) \nonumber \\
    &\times& K_n(q_{nb}, t_b | q_{na}, t_a; [x]) \rho_n(q_{na},q_{na}',t_a) .
    \label{Fn}
\end{eqnarray}
Here $K_n(q_{nb},t_b|q_{na},t_a; [x])$ is the wave function propagator for the $n$th harmonic oscillator, with boundary constraint $q_n(t_a) = q_{na}$ and $q_n(t_b) = q_{nb}$, and a functional dependence on the trajectory $x(t)$ of the central system, associated to the Lagrangian $L_n(q,q',x)$ in \eq{lagrangian2}.
The wave function propagator  $K$  of a harmonic oscillator can be calculated analytically for an arbitrary function $x(t)$, but to evaluate the corresponding influence functional (\ref{Fn}) also the initial density matrix of the  $n$th oscillator is needed.
On the analogy with the classical initial conditions (\ref{ic1}), the initial density matrix of the $n$th oscillator is here written as 
\begin{equation} \label{rhoc}
    \rho_n(q_{na},q_{na}',t_a) = \rho_\beta^{(n)}(q_{na}-X_a,q_{na}'-X_a) ,
\end{equation}
where $X_a$ represents the position of the central particle (see below) and $\rho_\beta^{(n)}(q,q')$ is the density matrix of a quantum harmonic oscillator 
with angular frequency $\oon$ in thermal equilibrium at an inverse temperature $\beta = 1 / \kB T$ \cite{Feynman1965a},
\begin{eqnarray}
	&&\rho_\beta^{(n)}(q,q') = 
	\label{rhoe}
	\\
	&&\kappa_n \exp \! \left\{ \! - \frac{m\oon}{2\hbar\sinh(\beta\hbar\oon)} 
			\! \left[
				\left( q^2 + q'^{\,2} \right) \cosh(\beta\hbar\oon) - 2 q q'
			\right]
		\! \right\}\! ,
		\nonumber
\end{eqnarray}
%
%
%
where the constant $\kappa_n=\sqrt{m\oon\rm{\tanh}(\beta\hbar{\mathit\oon}/2)/\pi\hbar}$
is fixed by the normalization condition   $\int d q \, \rho_\beta^{(n)}(q,q) = 1$.
As for the coordinate $X_a$, it can only depend on the initial coordinates $x_a$ and $x_a'$ of the central particle.
In general, the oscillator coordinates $q_{na}$ and $q_{na}'$  in \eq{rhoc} could have been shifted by different amounts $X_a$ and $X_a'$, writing the initial density matrix of the oscillator as $\rho_n(q_{na},q_{na}',t_a) = \rho_\beta^{(n)}(q_{na}-X_a,q_{na}'-X_a')$.
Since $\rho_n(q_{na},q_{na}',t_a)$ represents a state of thermal equilibrium,
it has to be invariant under a time reversal  -- equivalent to the exchange of $q_{na}$ and $q_{na}'$ in the oscillator density matrix,
which implies that $X_a = X_a'$ and that, at the same time,  $X_a$ is a symmetrical function of $x_a$ and $x_a'$.
Also, in a homogeneous environment, the density matrix in \eq{rhoc} has to be invariant under a spatial translation,
implying the same invariance for the expressions $q - X_a$ and $q' - X_a$.
Therefore $X_a$ is a symmetrical linear functions of $x_a$ and $x_a'$,
\begin{equation} \label{XA}
    X_a = \frac{x_a + x_a'}{2} .
\end{equation}
It is to be noticed that the form (\ref{rhoc}) of the initial conditions for the generic oscillator contains the adiabatic approximation in the quantum problem,
i.e. the assumption of thermal equilibrium for the environment degrees of freedom only.
A more rigorous treatment of the initial conditions would take into account the fact that the central particle is influenced
by the environment during the measurement which defines its initial state and 
all the degrees of freedom should be considered at the same time -- for a discussion of this point and more general forms of initial conditions 
see Refs.~\cite{Smith1987a,Schramm1987a,Grabert1988a}.
To summarize, in the initial conditions it is assumed (as in the classical model) that a fast relaxation of the environment around
the central particle has taken place and that the oscillator density matrix has a parametric dependence on the central particle coordinate.

Results can be conveniently re-expressed by introducing the coordinates
%
\begin{equation} \label{X}
    X = (x+x')/2 ,~~~~~~~~\xi = x' - x ,
\end{equation}
%
which, according to the interpretation of Schmid \cite{Schmid1982a}, represent the observable value of position and a measure of the corresponding quantum uncertainty, respectively.
Thus, \eq{XA} consistently represents the measured initial value of the central particle coordinate.
Also the result for the influence functional can be expressed through the new coordinates $X$ and $\xi$,
\begin{equation} \label{FS}
   F[X,\xi] = \exp\left(\frac{\iu}{\hbar} \Delta S[X,\xi]\right) ,
\end{equation}
where 
\begin{equation} \label{DS}
\Delta S[X,\xi] = 
                \int_{t_a}^{t_b} \! \! \! d t \! \! \int _{t_a}^{t} \! \! d s 
                \, \xi(t) [ u(t-s) \dot{X}(s)
              + \iu \alpha(t-s) \xi(s) ] ,
\end{equation}
with the same $u(t-s)$ defined by \eq{u} and 
%
\begin{equation} \label{alpha}
    \alpha(\tau) = 
    \sum_n\frac{m\hbar\oon^3}{2}\coth\left(\frac{\hbar\oon}{2\kB T}\right) \cos(\oon \tau) .
\end{equation}
%
The term linear in $\dot{X}$ in $\Delta S[X,\xi]$ corresponds 
to the classical dissipative force, while
the last term describes thermal fluctuations.
In the high temperature limit, that is when $\kB T \gg \hbar \Omega$, 
where $\Omega$ is a frequency cut-off of the frequency density $G(\oo)$,
one recovers the classical fluctuation-dissipation theorem, $\alpha(\tau) \to \kB T u(\tau)$.
Also the white-noise limit can be obtained, similarly to the classical case,  when $\Omega^{-1}$ is much smaller
than all the typical time scales of the system.
However, some care has to be used in this case,
since the requirement of a high $\Omega$ should be compatible with
the independent condition $\Omega \ll \kB T/\hbar$, 
which defines the high-temperature limit.

\subsection{Generalized quantum model}

In order to generalize the procedure illustrated above to a thermal environment moving
with a given velocity $v_{\rm env}$, one needs:\\
(a) the generalized Lagrangian of the total system; this has been already discussed in Sec.~\ref{gen-class};\\
(b) the quantum analogue of the classical equilibrium distribution \eq{ic3}\\
It is easy to verify that the density matrix $\rho_V(q,q')$ of a system of mass $m$
moving with average velocity $\v$ can be constructed 
from the corresponding density matrix $\rho_0(q,q')$ describing the system in its rest frame
by multiplying it by the factor $\exp[\iu m v_{\rm env} (q - q')/\hbar]$,
\begin{equation}
	 \rho_V(q,q') = \rho_0(q,q') \exp[\iu m v_{\rm env} (q - q')/\hbar] .
\end{equation}
This can be easily shown in the case of an isolated system with wave function $\phi_0(q)$ with a zero average momentum, $\bar{p}=0$,
which has a density matrix $\rho_0(q,q')=\phi_0(q)\phi_0^*(q')$ and probability density $P(q) = |\phi_0(q)|^2$. 
A system with wave function $\phi(q) = \phi_0(q) \exp(\iu m v_{\rm env} q/\hbar)$
has the same probability density $P(q)$ but an average momentum $\bar{p}=m v_{\rm env}$ and
a corresponding density matrix
$\rho_V(q,q') =\phi_0(q)\phi_0^*(q')\exp[\iu m v_{\rm env} (q-q')/\hbar] \equiv \rho_0(q,q')\exp[\iu m v_{\rm env}(q-q')/\hbar]$.
This is just what is needed here, since the probability density of the oscillator should remain unchanged (otherwise it would not represent anymore an equilibrium state) while only its average velocity is allowed to change.
Therefore, the initial density matrix of the $n$th oscillator is
\begin{eqnarray}
    && \rho_n(q_{na},q_{na}',t_a) = \nonumber \\
    && \rho_T^{(n)}(q_{na} \!-\! X_a, q_{na}' \!-\! X_a)
           \exp\!\!\left[\iu \frac{m v_{\rm env}}{\hbar} ( q_{na} \!-\!  q_{na}')\right],
           \label{ic3q}
\end{eqnarray}
where $\rho_T^{(n)}(q_{a}-X_a,q_{a}'-X_a)$ is the same density matrix defined 
by \eqs{rhoc}-(\ref{XA}).

The calculation of the influence functional now proceeds similarly to the case of the basic model,
but the new Lagrangian (\ref{lagrangian3}) and the new initial conditions (\ref{ic3q}) have to be used.
For the calculation of the oscillator wave function propagator  (which is defined by a quadratic action)
one only needs to know  the classical trajectory of the oscillators, defined by the equations of motion
$m\ddot{q}_n + \oon^2 q_n = \oon^2 (x + \dot{v}_\mr{env}/\omega^2)$,
in which the term $\oon^2 x$ is replaced now by $\oon^2 (x + \dot{v}_\mr{env}/\omega^2)$.
The same replacement can be done in the standard propagator,
$K_n(q_b,t_b|q_a,t_a; [x])$,  to compute the new propagator,
$K_n(q_b,t_b|q_a,t_a; [x+\dot{v}_\mr{env}/\omega^2])$.
The result is again of the form (\ref{FS}), where the effective action is now given by
\begin{eqnarray}
 && \Delta S[X,\xi] = \label{dS2} \\
 \nonumber
 && \frac{\iu}{\hbar} \int_{t_a}^{t_b} \! \! \! d t \! \int _{t_a}^{t} \! \! d s 
  \, \xi(t) \{ u(t-s) [ \dot{X}(s) - v_{\rm env}  ] + \iu \alpha(t-s) \xi(s) \} ,
\end{eqnarray}
in which the relative velocity $\dot{X}(s) - v_{\rm env}$ has replaced the central particle
velocity $\dot{X}(s)$ in \eq{DS}.

\section{Summary}
\label{summary}
%
The present paper has discussed the formulation of the Feynman-Vernon model of Brownian motion and introduced its generalization for a thermal environment moving with an assigned velocity $v_{\rm env}(t)$.
The formulation is summarized by \eq{ic3q}, which gives the initial conditions for the bath oscillators, and by the influence phase $\Delta S/\hbar$, defined by Eq.~(\ref{dS2}), describing the effective dynamics of the central particle.
The environment velocity $\v$ can be in turn a function of time, as long as it varies slowly enough not to change the state of thermal equilibrium of the oscillators.
The scheme presented above should be valid anyway for a constant environment velocity.
The model considered is expected to be suited for the study of problems involving the motion of particles in moving environments, such as superfluids.

\vspace{1cm}

\begin{acknowledgments}
This revision of Ref.~\cite{Patriarca-2005c} was made possible by the support of the European Regional Development Fund (ERDF) Center of Excellence (CoE) program grant TK133 and the Estonian Research Council through Institutional Research Funding Grants (IUT) No. IUT-39-1, IUT23-6, and Personal Research Funding Grant (PUT) No. PUT-1356. 
\end{acknowledgments}

\newpage


\onecolumngrid

\bibliography{fvmoving.bib}

\bibliographystyle{unsrt}

\end{document}